\documentclass{article}

\usepackage[utf8]{inputenc} 
\usepackage[T1]{fontenc}    
\usepackage{hyperref}       
\usepackage{url}            
\usepackage{booktabs}       
\usepackage{amsfonts}       
\usepackage{nicefrac}       
\usepackage{microtype}      
\usepackage{graphicx}
\usepackage{natbib}
\usepackage{doi}

\usepackage{arxiv}
\usepackage{graphicx}
\usepackage{subfigure}
\usepackage{setspace}
\usepackage{geometry} 
\usepackage{epstopdf}
\usepackage[labelsep=period]{caption}  
\usepackage[british]{babel} 
\usepackage[hang]{footmisc}
\usepackage{float}
\usepackage{amsmath}

\begin{document}

\title{Comparison of convolutional neural networks for cloudy optical images reconstruction from single or multitemporal joint SAR and optical images}
\date{}

\newcommand\splitcol[2]{
\begin{tabular}{@{}c@{}}#1 \\ #2\end{tabular}
}


\newcommand\loss{$l_1$}

\newcommand\meran{$SSOP_{mer}$}
\newcommand\merunet{$SSOP_{unet}$}
\newcommand\merunetdem{$SSOP_{unet+DEM}$}
\newcommand\merunetwosar{$SSOP_{unet,w/o SAR}$}
\newcommand\sca{$MSOP_{unet}$}
\newcommand\scadem{$MSOP_{unet+ DEM}$}
\newcommand\scawosar{$MSOP_{unet,w/o SAR}$}

\newcommand\dmer{SSOP}  
\newcommand\dcrga{MSOP} 
\newcommand\dgfa{MSOP\textsubscript{cld}} 

\author{ \href{https://orcid.org/0000-0003-4017-2635}{\hspace{1mm}Rémi Cresson} \\
    UMR TETIS, INRAE \\
	500 rue jean-francois breton, 34090 Montpellier \\
	\texttt{remi.cresson@inrae.fr} \\
	\And
	Nicolas Narçon \\
    UMR TETIS, INRAE \\
	\And
	Raffaele Gaetano \\
    UMR TETIS, CIRAD \\
	500 rue jean-francois breton, 34090 Montpellier \\
	\And
	Aurore Dupuis \\
    CNES \\
    18 Av. Edouard Belin, 31400 Toulouse \\
	\And
	Yannick Tanguy \\
    CNES \\
	\And
	Stéphane May \\
    CNES \\
	\And
	Benjamin Commandré \\
    UMR TETIS, INRAE \\
    
}

%

\begin{abstract}
With the increasing availability of optical and synthetic aperture radar (SAR) images thanks to the Sentinel constellation, and the explosion of deep learning, new methods have emerged in recent years to tackle the reconstruction of optical images that are impacted by clouds.
In this paper, we focus on the evaluation of convolutional neural networks that use jointly SAR and optical images to retrieve the missing contents in one single polluted optical image.
We propose a simple framework that ease the creation of datasets for the training of deep nets targeting optical image reconstruction, and for the validation of machine learning based or deterministic approaches.
These methods are quite different in terms of input images constraints, and comparing them is a problematic task not addressed in the literature.
We show how space partitioning data structures help to query samples in terms of cloud coverage, relative acquisition date, pixel validity and relative proximity between SAR and optical images.
We generate several datasets to compare the reconstructed images from networks that use a single pair of SAR and optical image, versus networks that use multiple pairs, and a traditional deterministic approach performing interpolation in temporal domain.
\end{abstract}

\keywords{Image reconstruction, clouds, atmospheric perturbation, SAR, optical, convolutional neural networks, gap-filling}

\maketitle


\section{Introduction}
\label{s_intro}
 
\sloppy

\subsection{Context}

The Sentinel constellation is composed of different coupled SAR and optical sensors with short revisit period (five to ten days).
However, optical images are frequently polluted by cloud cover.
To leverage the problem of optical image reconstruction, various approaches have been proposed over the years.
First, approaches based on mathematical, physical or statistical model, have been extensively used to reconstruct the missing parts of the images. 
A review of these traditional approaches are summarized in \cite{shen2015missing}.
Among them, we can distinguish multispectral based~\cite{hu2015thin} methods, multitemporal~\cite{cheng2014cloud}~\cite{li2014recovering}, and methods using optical and SAR data fusion~\cite{eckardt2013removal}.
Lastly, machine learning and particularly deep learning have become popular to achieve the task of cloudy images reconstruction, thanks to the unprecedented ability to fuse images of different modalities, and accompanied with state of the art results.
In recent years, deep neural networks have proven to be effective for image reconstruction from time series of same modality~\cite{zhang2018missing} or from timely available images at coarser spatial resolution~\cite{liu2019stfnet}, or from joint optical and SAR time series~\cite{scarpa2018cnn}\cite{cresson2019optical}.
\cite{sarukkai2020cloud} have casted the problem of cloud removal as a conditional image synthesis challenge, and have proposed a network to remove clouds from a single optical image or from a triplet of optical images.
In \cite{meraner2020cloud}, a cloudy optical image is reconstructed with the help of a single SAR, using a convolutional neural network with a residual correction performed on the input cloudy optical image.
Conversely, \cite{ebel2020multisensor} have jointly modeled the cloud removal and the synthetic cloudy optical images generation problems, concluding that the networks trained over real data were performing the best.

\subsection{Problematic}

Our purpose is to lead the evaluation of several approaches based on convolutional neural networks trained on real data, that reconstruct optical images impacted by clouds. 
We compare these approaches with a popular traditional deterministic approach, the gap-filling \cite{inglada_gapfil}.
While the existing literature covers mostly the comparison of approaches that consume the same inputs, comparing approaches consuming various forms of inputs, e.g. single or multiple, optical and/or SAR images or pairs of images, remains an interesting topic from an operational perspective.
In this paper, we address the comparison of the following kinds of methods, which consume one or more input pair of optical and SAR images acquired in various conditions to reconstruct or generate one single output optical image:
\begin{enumerate}
\item Reconstruct a cloudy optical image using an additional SAR image acquired at the same date,
\item Reconstruct a cloudy optical image using an additional SAR image acquired at the same date, and two other cloudy optical/SAR images pairs acquired before and after,
\item Generate an optical image at one given desired date using two clean optical images acquired before and after.
\end{enumerate}

\subsection{Method}

In this paper, we provide an insight into various optical image reconstruction methods.
In particular, we address the question of which approach to employ for a specific availability of remote sensing products, in comparing a few selected single date based and multiple dates based methods.
Since these methods employ inputs of different nature (single optical or SAR image, or optical and SAR images pair), number (single or multiple image or pair or images), and cloud coverage (clean or cloudy images), their comparison is not straightforward.
To leverage this, we introduce the \textit{acquisitions layout}, a descriptor of the available inputs and their properties, for a specific approach.
We then use space partitioning data structures to ease the generation of various datasets from specific acquisitions layouts.
These datasets are then used to train networks, and also at inference time for the comparison of the different methods when a common set of inputs can be shared and matches the expected constraints, e.g. cloud coverage.
This simple yet generic framework allows to produce datasets tailored for a specific problem and suited to the data availability, i.e. inputs and targets images.
We carry out the benchmarks of representative state of the art methods for optical image reconstruction, namely the network presented in \cite{meraner2020cloud}, which uses a single pair of optical and SAR image acquired near the same date, and a convolutional network that inputs three consecutive pairs to reconstruct the central optical image \cite{cresson2019optical}\cite{scarpa2018cnn}.
In \cite{scarpa2018cnn}, an additional DEM is used as input of the network, and we also investigate the contribution of such ancillary data in the single date network.
To better disentangle the benefits of the different modalities, we perform an ablation study removing the DEM, and the SAR inputs.

\subsection{Overview}

In section \ref{s_data}, we present the remote sensing data used in this study.
In section \ref{s_architectures}, we detail the implemented models.
In section \ref{s_datasets}, we detail our framework for the creation of datasets, which is a crucial aspect of our work.
In section \ref{s_benchmarks} we detail the methodology used to train the models and carry out the comparison of the different approaches.
Finally we discuss the results in section \ref{s_discussion}.

\section{Data}
\label{s_data}

\subsection{Sentinel images}

\begin{figure}[ht!]
\begin{center}
\includegraphics[width=1.0\columnwidth]{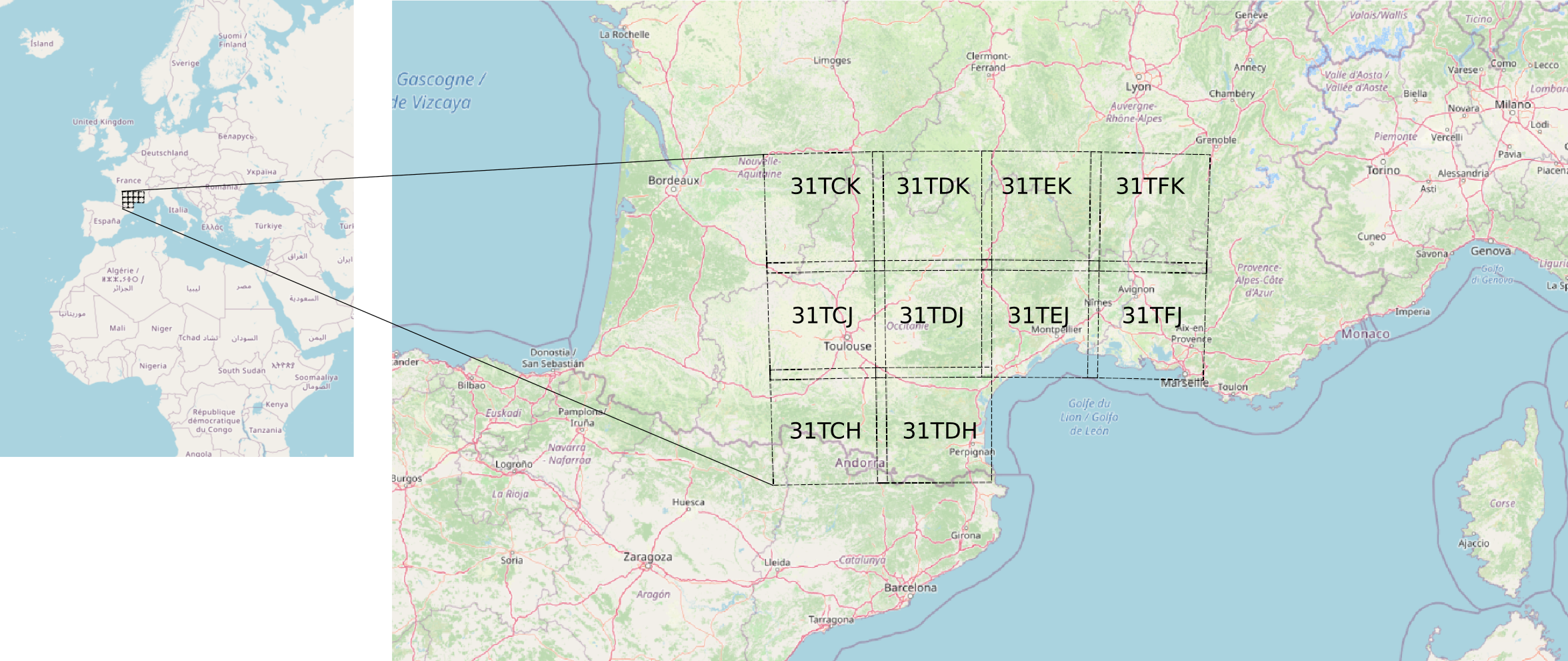}
\caption{The region of interest, located in the Occitanie area (south of France mainland). Sentinel-2 images envelopes are plotted in black. Map data © OpenStreetMap contributors, CC BY-SA}
\label{fig_roi}
\end{center}
\end{figure}

We use 10 tiles of Sentinel-2 images acquired over the Occitanie area in France (figure \ref{fig_roi}), from january 2017 to january 2021, that represents a total of $3593$ optical images.
We also use every available Sentinel-1 images acquired in ascending orbit over the Occitanie area during the same period, that we superimpose over the Sentinel-2 images pixels grids (more details are provided in section \ref{sss_s1}), which represents a total of $5136$ SAR images.
We believe that since a large part of the earth is covered only with single orbit (i.e. ascending or descending), our study results would be more easily reproducible with a single orbit for SAR images, hence we use only the ascending orbit over our study area.
Table \ref{tab_nimages} summarizes the number of Sentinel images used for this study.
The total area covered by the Sentinel tiles is $106.7\times10^3 km^2$.
The following sections details the Sentinel-1 and Sentinel-2 products.

\subsubsection{Sentinel-1 images}
\label{sss_s1}

We have used the so-called S1Tiling tool\footnote{https://gitlab.orfeo-toolbox.org/s1-tiling/s1tiling} to automatically download and process the Sentinel-1 images.
The tool performs the orthorectification and the calibration in sigma nought of the VV and VH SAR images channels.
It also projects and resamples the final images over the same coordinate reference system and pixel grid as the Sentinel-2 images, at 10m physical spacing resolution.

\subsubsection{Sentinel-2 images}

\begin{figure}[ht!]
\begin{center}
\includegraphics[width=1.0\columnwidth]{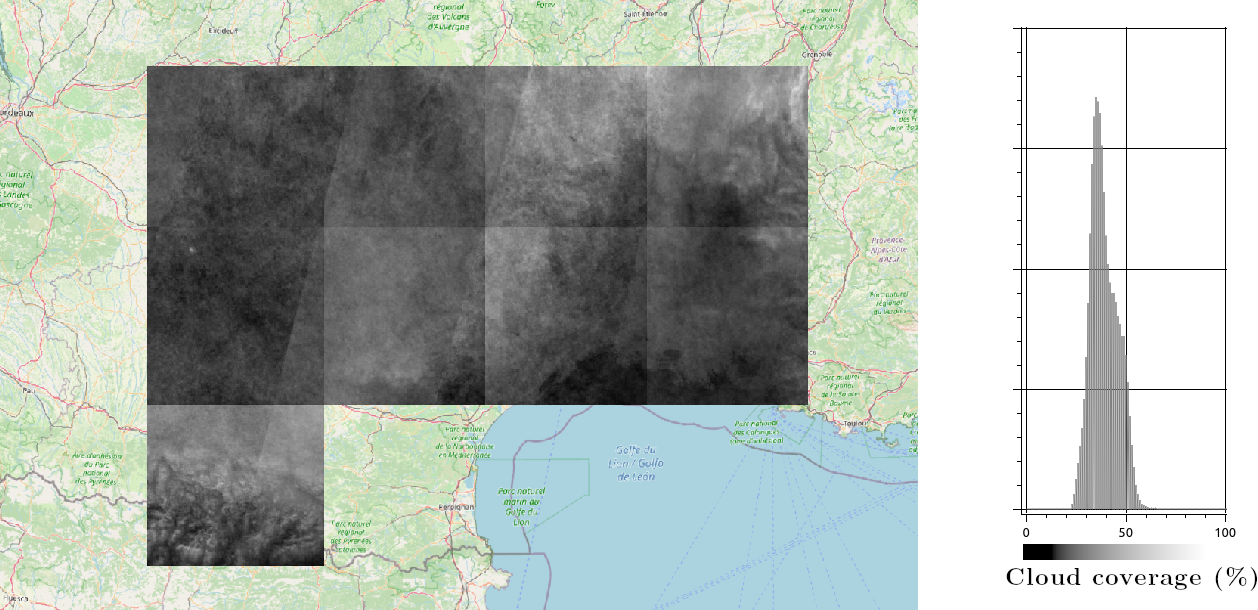}
\caption{The cloud coverage over the area. Left: the percentage of cloud-free pixels in computed at each location. Right: histogram of the values. Mean cloud coverage: $39.13\%$, standard deviation: $6.80\%$. Map data © OpenStreetMap contributors, CC BY-SA}
\label{fig_cloudcoverage}
\end{center}
\end{figure}

The Theia Land data center\footnote{https://www.theia-land.fr/en/product/sentinel-2-surface-reflectance/} provides Sentinel-2 images in surface reflectance.
The products are computed using MACCS (Multi-sensor Atmospheric Correction and Cloud Screening), a level 2A processor which detects the clouds and their shadows, and estimates aerosol optical thickness, water vapour and corrects for the atmospheric effects \cite{hagolle2015spot}.
While level 1C processing level could have been used in this study, we chose level 2A products because they include cloud masks that are useful meta-data. 
Indeed these information suffice to derive a cloud coverage percentage over patches.
Figure \ref{fig_cloudcoverage} shows the cloud coverage computed for each location from the number of cloudy pixels among available ones in the temporal dimension.
The average cloud coverage over the area is $39.1\%$ and the standard deviation $6.8\%$.
An evaluation of the cloud masks is provided in \cite{baetens2019validation}.
To discharge storage and computational requirements, we used only the 10m spacing bands, i.e. spectral bands number 2, 3, 4 and 8.

\begin{table}[h]
\begin{center}
\begin{tabular}{|l|c|c|}
 \hline
 Tile & S1 & S2 \\
 \hline
31TCH & 684 & 374 \\
31TCJ & 456 & 380 \\
31TCK & 454 & 273 \\
31TDH & 462 & 370 \\
31TDJ & 462 & 302 \\
31TDK & 462 & 286 \\
31TEJ & 672 & 512 \\
31TEK & 595 & 311 \\
31TFJ & 443 & 408 \\
31TFK & 446 & 377 \\
 \hline
\end{tabular}
\caption{Number of Sentinel-1 (S1) and Sentinel-2 (S2) images used over the Occitanie area (France)}
\label{tab_nimages}
\end{center}
\end{table}

\subsubsection{Digital Elevation Model}

The Digital Elevation Model (DEM) from the Shuttle Radar Topography Mission (SRTM)\cite{farr2000shuttle} is used to perform the processing of Sentinel-1 images.
It is also used in our \merunetdem{} and \scadem{} networks, introduced in the following sections.

\section{Architectures}
\label{s_architectures}

The implemented deep learning based models are detailed in the subsections below.
The number of trainable parameters, batch size used per GPU, and training time is summarized in table \ref{tab_modelssize} for each model.

\begin{table}[ht]
\begin{center}
\begin{tabular}{|c|c|c|c|}
\hline
Model    & Nb. params & Batch & Training time \\ \hline
\meran   & 18,905M & 8  & 3136h.GPU \\ \hline
\merunet & 16.652M & 64 & 104h.GPU \\ \hline
\sca     & 29.272M & 32 & 149h.GPU\\ \hline
\end{tabular}
\caption{Number of parameters in the trained models, batch size used, and training time (in hours.GPUs) for the training step on a NVIDIA V100 GPU with 32Gb RAM. The training time corresponds to the duration required to computed the best model.}
\label{tab_modelssize}
\end{center}
\end{table}

\subsection{Single SAR/Optical pair (SSOP)}

\begin{figure}[H]
\begin{center}
\includegraphics[scale=0.4]{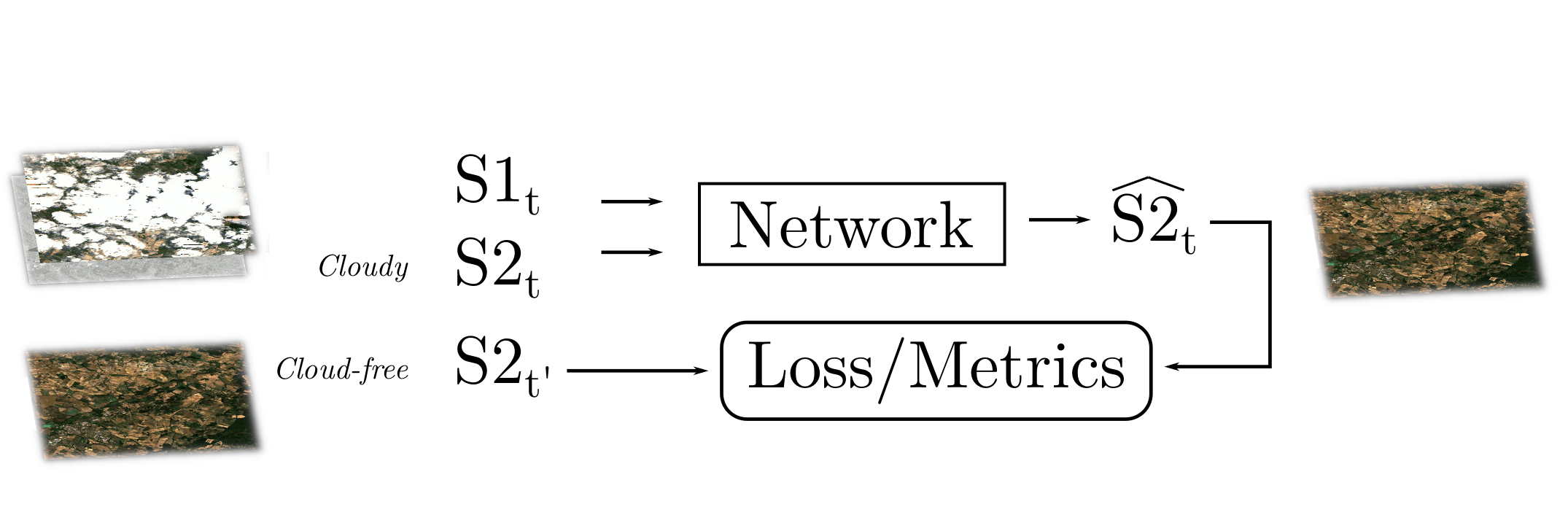}
\caption{SSOP network. $S1_t$ and $S2_t$ denotes the input pair of Optical and SAR images acquired at date $t$. The $S2_{t'}$ denotes the reference image used to compute the loss (during training) and the metrics (at inference time from test data) from the reconstructed optical image $\widehat{S2_t}$.}
\label{fig_meraner}
\end{center}
\end{figure}

We denote \textit{SSOP}, the approaches that input one SAR image and one optical image polluted by cloud, and which reconstruct the missing parts of the optical image. 
Introduced in \cite{meraner2020cloud}, this kind of approach is trained and evaluated from samples composed of $(S1_t$, $S2_t$, $S2_{t'})$ triplets, where $S2_t$ is an optical image potentially polluted by clouds, with $S1_t$ and $S2_t$ acquired close together and $S2_{t'}$ a cloud-free optical image acquired close to the ($S1_t, S2_t$) pair.
Figure \ref{fig_meraner} illustrates the architecture of this family of networks.

\subsubsection{\meran}

We implement the network described in \cite{meraner2020cloud}, which uses a residual correction from a ResNet backbone \cite{he2016deep} to reconstruct the output optical image from a pair of one SAR image and one cloudy optical image. 
However, the authors of \cite{meraner2020cloud} did not have $S2_{t'}$ acquisitions systematically close to $S2_t$, therefore they use an additional loss based on cloud masks to encourage the identity transformation of cloud-free pixels.
Since our goal is to use a simple information about the presence of clouds, namely an approximation of the cloud cover percentage in a Sentinel-2 image or a set of patches, we don't use such pixel-wise cloud-mask based loss.
Moreover, as we control the gathering of samples matching the acquisitions layout described in table \ref{tab_al_meraner}, a cloud-free $S2_{t'}$ image acquired close to $S2_t$ is always available.
We hypothesize that the contribution of such loss is likely marginal since none or very little changes should happen between $S2_{t'}$ and $S2_t$. 
Hence we train the network using only the \loss{} loss.
We denote \meran{} the implementation of this network.

\subsubsection{\merunet}

\begin{figure}[ht!]
\begin{center}
\includegraphics[width=1.0\columnwidth]{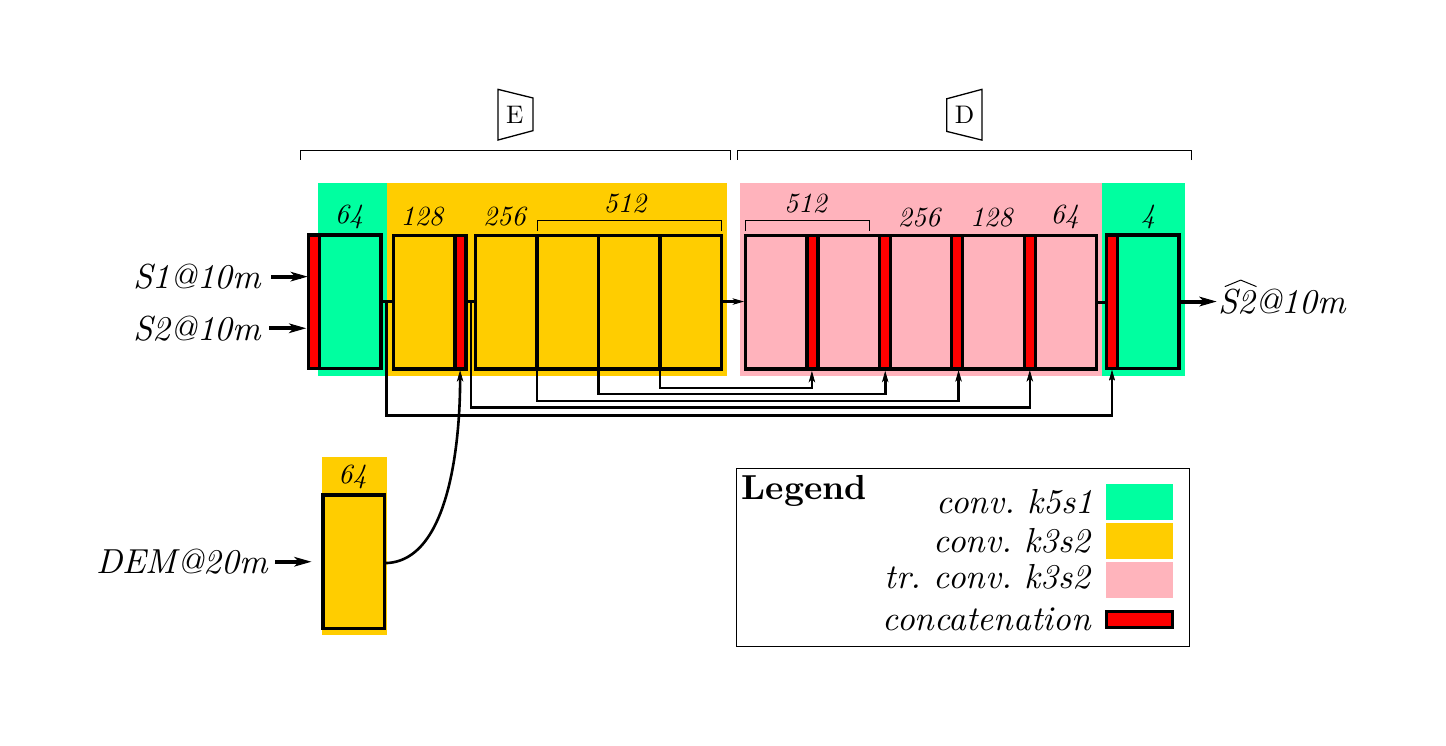}
\caption{Proposed architecture for encoder ($E$) and decoder ($D$), enabling the use of inputs of different resolutions. First and last convolutions use a unitary stride and a kernel of size 5. Other convolutions use strides 2 and a kernel of size 3. Skip connections between the encoder and the decoder perform the concatenation of the features from the encoder with the decoder outputs. All convolutions except the last \textit{k5s1} convolution are followed with a ReLU activation function. No batch normalization is used.}
\label{fig_unet}
\end{center}
\end{figure}

\begin{figure}[H]
\begin{center}
\includegraphics[scale=0.4]{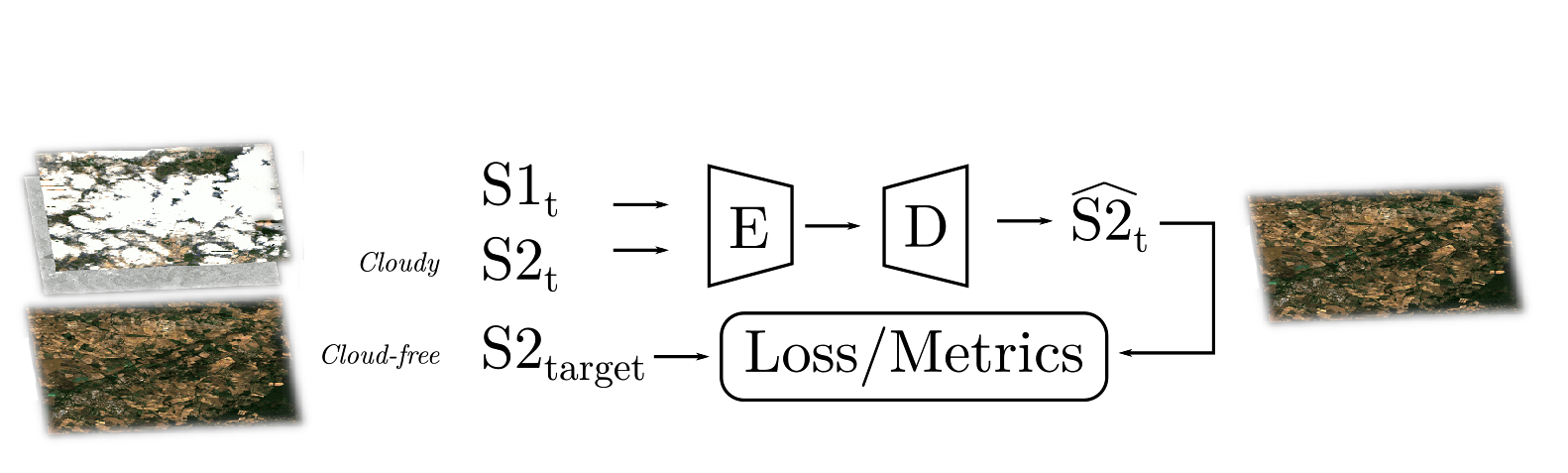}
\caption{Our modified SSOP network. $S1_t$ and $S2_t$ denotes the input pair of Optical and SAR images. The $S2_{t'}$ denotes the reference image used to compute the loss (during training) and the metrics (at inference time from test data) from the reconstructed optical image $\widehat{S2_t}$. (E) and (D) denotes respectively the encoder and the decoder of the U-Net backbone.}
\label{fig_merunet}
\end{center}
\end{figure}

We implement a modified version of the previously described architecture, employing a U-Net backbone \cite{ronneberger2015u} instead of ResNet. 
Our motivation behind this modification is that the ResNet backbone has two disadvantages compared to U-Net: 
(i) convolutions are applied without any stride, which consumes a lot more memory and requires much more operations since all convolutions apply over the entire images at their native resolution i.e. without any downsampling of the features, 
(ii) all inputs have to be resampled at the smallest input images resolution, i.e. a physical spacing of 10 m, since all the network features are computed at the same resolution and no downscaling is performed across the network, which is computationally counterproductive.
We denote \merunet{} our implementation of this modified network.
To illustrate the advantage of additional inputs at a lower resolution, we use an additional Digital Elevation Model (DEM) as input, resampled at 20m, as shown in figure \ref{fig_unet}.
Our modified network is illustrated in figure \ref{fig_merunet}.
We denote \merunetdem{} the model with the input DEM.
It can be noted that such a model could also generate outputs of different resolutions, typically the Sentinel-2 spectral bands at 20m.

\subsection{Multiple SAR/Optical pairs (MSOP)}

A number of approaches using multiple pairs of optical and SAR images have been presented in the literature.
For instance, \cite{scarpa2018cnn} use two pairs of SAR/optical acquired before and after date $t$, a Digital Elevation model (DEM), and an additional SAR image acquired at date $t$ to estimate radiometric indices at date $t$.
Conversely, \cite{cresson2019optical} use multiple optical and SAR images to generate a synthetic optical image at date $t$.
While these works were carried on input cloud-free images, a similar network architecture can also be applied on cloudy input images to retrieve the missing contents of the optical image at date $t$.

\subsubsection{\sca}

\begin{figure}[H]
\begin{center}
\includegraphics[scale=0.4]{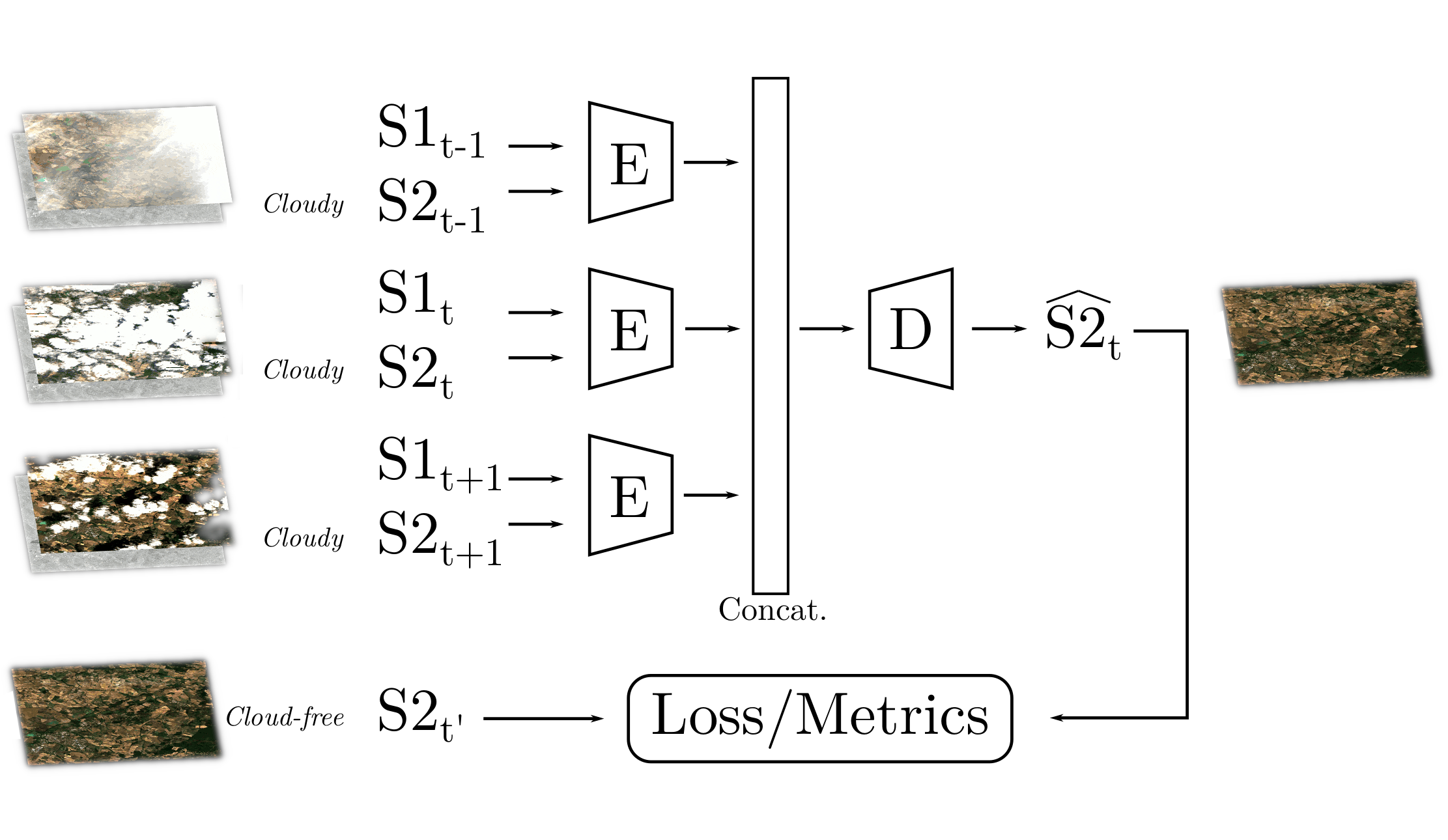}
\caption{\sca{} network. $(S1_{t-1}, S2_{t-1})$, $(S1_t, S2_t)$ and $(S1_{t+1}, S2_{t+1})$ denotes the input pairs of optical and SAR images. $S2_{t'}$ denotes the reference image used to compute the loss (during training) and the metrics (at inference time from test data) from the reconstructed optical image $\widehat{S2_t}$.}
\label{fig_crga_os2}
\end{center}
\end{figure}

We build a multitemporal network inspired from the architectures presented in \cite{scarpa2018cnn} and \cite{cresson2019optical}.
We generalize to multitemporal the approach of \cite{meraner2020cloud} with a new architecture that inputs mutliple SAR/optical images pairs at $t-1$, $t$ and $t+1$ and a DEM, aiming to reconstruct the potentially damaged optical image at $t$.
We use a similar architecture as the encoder/decoder U-Net backbone of the \merunet{} model, except that encoder weights are shared for $t-1$, $t$ and $t+1$ inputs, and features from three encoders ($E$) are concatenated before being processed with the decoder ($D$), which outputs the reconstructed optical image at $t$.
Unlike \cite{meraner2020cloud}, our model does not employ residual connections to generate the reconstructed optical image (figure \ref{fig_crga_os2}).

We denote \scadem{} the MSOP model using the input DEM in (E) as shown in figure \ref{fig_unet}.

\subsection{Gap-filling}

The \textit{Gap-filling} consists in interpolating temporally close optical images to approximate one target image \cite{inglada_gapfil}.
While gap-filling is not a reconstruction method, i.e. the input image at $t$ is not used, it is commonly used as such, in estimating the parts of the image that is polluted by clouds.
Gap-filling is restricted to cloud-free input images, and do not use SAR images.
In the case of a linear model, the generated output image $\widehat{S2_t}$ can be written using the following formula:
\begin{equation}
\widehat{S2_t} = S2_{t-1} + (S2_{t+1} - S2_{t-1}) \times \frac{T_{t} - T_{t-1}}{T_{t+1} - T_{t-1}}
\end{equation}
Where $T$ is the timestamp, in seconds, of the dates.

\section{Datasets}
\label{s_datasets}

\subsection{Acquisitions layouts}
\label{ss_als}

The so-called \textit{Acquisitions layout} describes inputs and targets of a specific use-case scenario.
In the particular case of image reconstruction addressed in this paper, the acquisitions layouts presented in the following sections have one common item, namely $S2_{t'}$ the target cloud-free optical image.
Depending on the approaches, the acquisitions layout can include additional items, for instance:
\begin{itemize}
\item A single SAR image
\item A single optical image
\item A pair of SAR + optical image
\end{itemize}

For each item, the acquisitions layout describes crucial properties:
\begin{itemize}
\item For each optical image: a range of cloud coverage percentage (e.g. $[0, 10]$),
\item For each (SAR, optical) pair: the maximum temporal gap between the two images acquisition dates, in hours,
\item For each SAR or optical image: the acquisition date range, relative to a reference item of the acquisitions layout (e.g. $[240h, 360h]$).
\end{itemize}

\begin{figure}[ht!]
\begin{center}
\includegraphics[width=0.7\columnwidth]{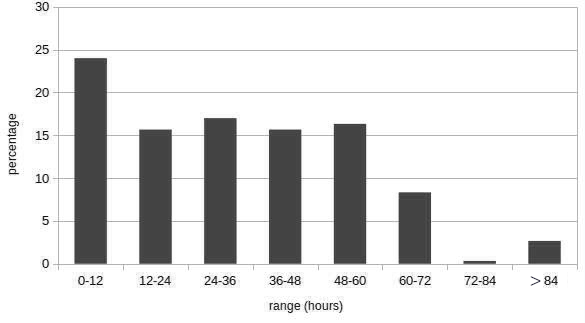}
\caption{Distribution of the temporal gap between the closest (S1, S2) images from the available images.}
\label{fig_hist}
\end{center}
\end{figure}

This simple yet generic description formalizes of how the images are acquired for a particular use-case scenario.
We have carefully crafted acquisitions layouts that represents the operational context of use of the approaches, i.e. for which it is possible to use them on every available images.
For instance, to chose the maximum temporal gap between the SAR and the optical images acquisition dates, we have analyzed the distribution of the temporal gap between the closest (S1, S2) images (figure \ref{fig_hist}).
Since more than $96\%$ of the nearest (S1, S2) pairs are close to $72$ hours, we used this duration as the maximum temporal gap in SAR-optical images pairs.
We provide in section \ref{s_datasets} all acquisitions layouts suited for the training and testing of the involved networks, and explain how the other parameters (i.e. time ranges for each acquisition layout items) are chosen.

\subsection{Patches indexation}

\begin{figure}[ht!]
\begin{center}
\includegraphics[width=1.0\columnwidth]{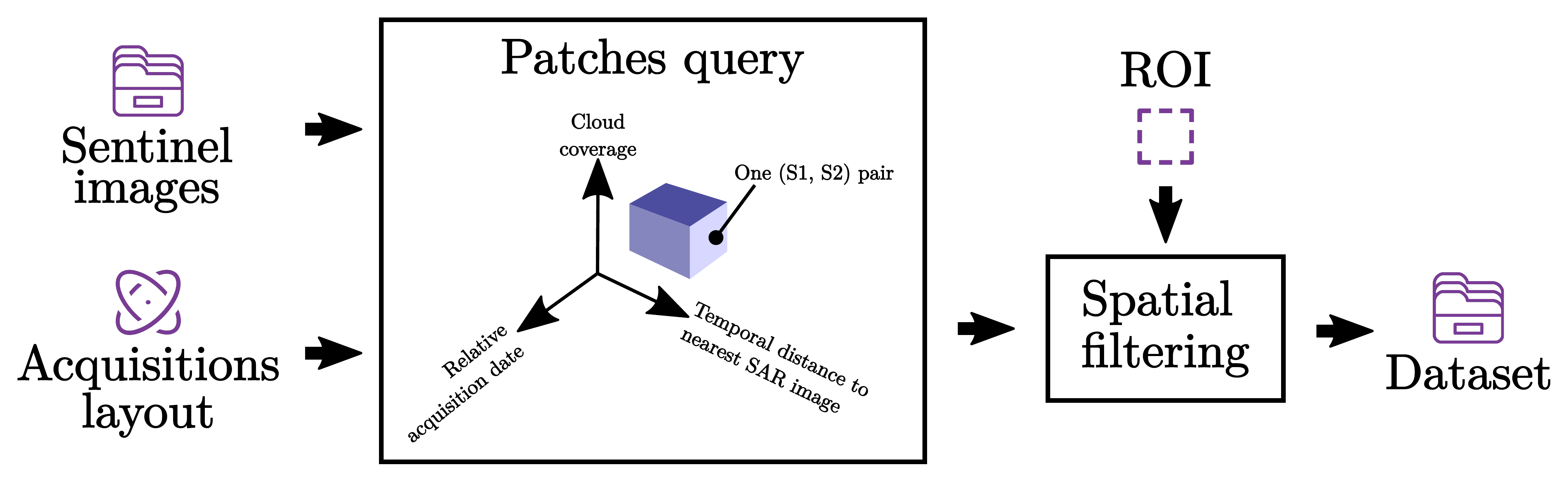}
\caption{Workflow for the creation of datasets. The sample query use a R-Tree indexing the available Sentinel-1 and Sentinel-2 images patches.}
\label{fig_workflow}
\end{center}
\end{figure}

Figure \ref{fig_workflow} gives an overview of the dataset creation step.
In order to perform the query of samples, i.e. the search of groups of patches that match the properties defined in the acquisitions layout, we use an indexation structure.
An R-Tree indexes all available patches.
The space partitioning data structure describes the following dimensions:
\begin{itemize}
\item Cloud coverage (ranging from $0\%$ to $100\%$)
\item Time from the reference image of the acquisitions layout
\item Duration to the closest valid SAR patch
\item Number of pixels different from the no-data value
\end{itemize}
A Kd-Tree is used to ease the computation of the duration to the closest valid SAR patch.
The R-Tree is built for each patches of Sentinel tiles using the acquisition dates provided in the Sentinel images metadata, and the following statistics collected on the Sentinel images patches:
\begin{itemize}
\item \textbf{For Sentinel-1 images}: the number of valid pixels,
\item \textbf{For Sentinel-2 images}: the number of valid pixels, and the number of pixels impacted by clouds. To compute this last, we use the cloud quality mask provided in the Theia product.
\end{itemize}

We note that cloud masks are only used as a single value for each patch, representing the proportion of cloud coverage.
We have computed the R-Trees from non-overlapping, $256\times256$ sized patches in all Sentinel tiles.
This indexation structure is computed once.
After that, any acquisitions layout can be used to query all samples matching the defined properties of the remote sensing acquisitions.
One generated sample includes data arrays containing pixels and ancillary data, e.g. acquisition date, for each items of the acquisitions layout.
Finally, the samples are restricted in the provided region of interest, to allow the generation of mutually exclusive samples in the geographical domain, i.e. in training, validation and test datasets.

\subsection{Generation of samples}

For each network, samples are first extracted in the images from the specific acquisitions layout.
We split the samples in three groups: training, validation, and test.
We ensure that these three groups are mutually exclusive by randomly selecting their location in the geographical domain, without overlapping patches of distinct groups.
We have randomly selected $5\%$ and $15\%$ of the area to form the region of interest for the validation and test datasets, and the other $80\%$ has been used to form training datasets.
Since the swath of Sentinel-1 and 2 does overlap in some areas, the samples density is heterogeneous in spatial domain.
For this reason, we have limited the number of samples per spatial location in the training and validation datasets, to guarantee that models are trained with the same number of patches at each location.
Thus for the training and validation datasets, a maximum amount of $50$ samples has been collected at each locations.
For the test dataset, all available samples are extracted.
Table \ref{tab_nsamples} summarize the number of distinct samples for training, validation and test datasets.
The differences in samples number is due to the availability of images, or pairs of images, depending on the properties defined in the acquisitions layout, i.e. number and type of acquisitions, cloud coverage and temporal constraints.
All acquisitions layouts used to generate the datasets are detailed in the following subsections.

\begin{table}[H]
\begin{center}
\begin{tabular}{|l|c|c|c|}
\hline
Dataset & Training & Validation & Test \\
\hline
\dmer  & $600.1k$ & $35.3k$ & $70.6k$ \\
\dcrga & $600.1k$ & $35.3k$ & $70.6k$ \\ 
\dgfa  & /        &   /     & $76k$   \\
\hline
\end{tabular}
\caption{Number of samples in each datasets. The \dmer{}, \dcrga{} and \dgfa{} datasets are used respectively to train the mono-temporal networks, the multi-temporal networks, and to compare all models with the gap-filling over an acquisitions layout matching all approaches validity domains.}
\label{tab_nsamples}
\end{center}
\end{table}

In the following sections, we detail the properties of each acquisitions layout of the datasets.

\subsubsection{\dmer{} (Single date inputs cloudy optical image)}  
\label{sss_al_meraner}

The acquisitions layout for SSOP networks training is presented in table~\ref{tab_al_meraner}.
We chose the following parameters for the acquisitions layout: the maximum gap between the acquisition dates of S1 and S2 images is set to $72$ hours and the maximum spread between the cloud-free and the polluted optical images is set to $10$ days. 
With this settings, we reach a total number of $600k$ samples for training, which is approximately 4 times the amount of samples used in the original paper of \cite{meraner2020cloud}.

\begin{table}[H]
\begin{center}
\begin{tabular}{l|c|c|c}
\hline
Name  & S1 ($\pm \Delta t$) & S2 (\% clouds) & Time-stamp   \\
\hline
$t$   & Yes ($\pm72$h)      & $[0, 100]$     & Reference    \\
$t'$  & /                   & $0$            & $[-10d, +10d]$ \\
\hline
\end{tabular}
\caption{Acquisitions layout for the SSOP networks, depicting the ($S1_t$, $S2_t$) images pair acquired close to a single cloud-free optical image ($S2_{t'}$).}
\label{tab_al_meraner}
\end{center}
\end{table}

\subsubsection{\dcrga{} (Multitemporal inputs, any optical images i.e. cloudy or not)} 
\label{sss_al_os21}

The acquisitions layout for the training of the \sca{} network is presented in table \ref{tab_alcrga_os21}.
It consists of three optical images at $t-1$, $t$ and $t+1$ that can be polluted by clouds, and one cloud-free optical image at $t'$, used as the training target.
As explained in section \ref{ss_als}, we used $72$ hours for the maximum gap between the acquisition dates of S1 and S2 images.
The cloud-free optical image is acquired at most $10$ days from the optical image at $t$, to roughly falls within the less frequent revisit cycle of the Sentinel-2 constellation everywhere over our study area.
Finally, we have selected a temporal range for $t-1$ and $t+1$ dates that avoids the cloud-free optical image acquisition date, and that also falls within the revisit cycle of the Sentinel-2 constellation, i.e. $10$ to $18$ days.

\begin{table}[H]
\begin{center}
\begin{tabular}{l|c|c|c}
 \hline
 Name    & S1 ($\pm \Delta t$) & S2 (\% clouds) & Time-stamp     \\
 \hline
$t-1$    & Yes ($\pm72$h)     & $[0,100]$       & $[-18d, -10d]$ \\
$t$      & Yes ($\pm72$h)     & $[0,100]$       & reference      \\
$t'$     & /                  & $0$             & $[-10d, +10d]$ \\
$t+1$    & Yes ($\pm72$h)     & $[0,100]$       & $[+10d, +18d]$ \\
 \hline
\end{tabular}
\caption{Acquisitions layout used to train the \sca{} network.}
\label{tab_alcrga_os21}
\end{center}
\end{table}

\subsection{\dgfa{} (Multitemporal inputs, with cloud-free optical images at $t-1$ and $t+1$, and one cloudy image at $t$)} 
\label{sss_al_crga_gf}

Table \ref{tab_crga_gfa} shows one acquisitions layout enabling the comparison of the gap-filling with the SSOP and MSOP networks, thanks to cloud-free optical images available at $t-1$ and $t+1$, and one completely cloudy optical images at $t$, which intends to make as fair as possible the comparison.
We denote the corresponding dataset \dgfa{}.
In this acquisitions layout, the cloud-free optical image at $t'$, acquired at most $5$ days from the date $t$, is used to compute the metrics over the reconstructed image.

We note that our settings make possible the extraction of a sufficient number of samples, thanks to the availability of Sentinel-1 and Sentinel-2 over our study site, but this setting might be adjusted for other regions of the world where the Sentinel coverage is less timely available.

\begin{table}
\begin{center}
\begin{tabular}{l|c|c|c}
 \hline
 Name    & S1 ($\pm \Delta t$) & S2 (\% clouds) & Time-stamp     \\
 \hline
$t-1$    & Yes ($\pm72$h)     & $0$             & $[-18d, -10d]$ \\
$t$      & Yes ($\pm72$h)     & $100$           & reference      \\
$t'$     & /                  & $0$             & $[-5d, +5d]$   \\
$t+1$    & Yes ($\pm72$h)     & $0$             & $[+10d, +18d]$ \\
 \hline
\end{tabular}
\caption{Acquisitions layout enabling the comparison of the gapfilling and the \sca{} network. The cloud coverage at $t$ is $100\%$, all other optical images remain cloud free.}
\label{tab_crga_gfa}
\end{center}
\end{table}

\section{Benchmarks}
\label{s_benchmarks}

\subsection{Protocol}

We train all networks with their respective datasets presented in section \ref{s_data}.
The SSOP and MSOP models are trained over the training dataset detailed in section \ref{sss_al_meraner}, and evaluated over the test datasets detailed in sections \ref{sss_al_os21} and \ref{sss_al_crga_gf}.
The MSOP models are trained over the dataset detailed in section \ref{sss_al_os21}, and evaluated on the test datasets detailed in sections \ref{sss_al_os21} and \ref{sss_al_crga_gf}.
We use the ADAM algorithm \cite{kingma2014adam} to train all networks to minimize the \loss{} loss.
For \sca{} and \merunet{}, we use a learning rate of $lr=0.00012$, $\beta_0=0.9$ and $\beta_1=0.999$, with a batch of size 128 distributed across 4 GPUs.
We train the \meran{} network using the same setup as described by the authors.
All experiments are realized on NVIDIA V100 GPUs with 32Gb RAM.
We kept the trained models that reach the smallest \loss{} loss on the validation split.
To assess the performance of each approach, we compute the following metrics between the reconstructed output optical images $\widehat{S2_t}$ and the reference cloud-free optical image $S2_{t'}$ over the test datasets:
\begin{itemize}
\item Peak Signal to Noise Ratio (PSNR):
\begin{equation}
PSNR = 10 \times log_{10}(\frac{d^2}{MSE})
\end{equation}
Where MSE is the Mean Squared Error computed over $n$ patches:
\begin{equation}
MSE=\frac{1}{n}\sum^n{\lVert \widehat{S2_t} - S2_{t'} \rVert^2}
\end{equation}
The higher is the PSNR, the closer are the values of the estimated image to the target image.
\item Spectral angle (SAM)\cite{kruse1993spectral}, representing the mean spectral angle between the estimated image and the target image, ranging in $[0, \pi]$
\item The Structural Similarity Index (SSIM)\cite{wang2004image}, measures the similarity in terms of structure, between the estimated image and the target image. The range is $[0, 1]$, and values close to $1$ correspond to the best structural similarity between the images.
\end{itemize} 

\subsection{Ablation study}

In order to assess the SAR, DEM, and optical modalities benefits, we have performed an ablation study.
For the SSOP and MSOP networks, we have derived two modified architectures, one without the SAR input, and the other without DEM and without SAR.
For a sake of computational budget, we only have studied the ablation with the U-Net based networks.

\subsection{Results}

In the following section, we report the evaluation metrics computed on all test datasets. 

\subsubsection{Comparison of SSOP networks}
\label{sss_res_ssop}

The comparison between SSOP networks is carried out on the test dataset presented in section \ref{sss_al_meraner}.
We first report the metrics obtained with SSOP networks in table \ref{tab_results_ssop}.
It can be noticed that the metrics from the modified network (\merunet{}) are close to the original \meran{}.
SSIM and SAM are slightly better for \meran{} and MSE and PSNR a bit better for \merunet{}. 
While this result is not groundbreaking in terms of evaluation metrics, we highlight the huge difference of required overall processing time: to train both networks over the same dataset with the same setup, \merunet{} needs $\approx 30$ hours and \meran{} $\approx 35$ days.
For this particular reason, we have chosen to perform all other benchmarks only on \merunet{}, since the metrics are quite similar to \meran{}, but the processing budget far lower and we could ran more experiment at lower cost.
It can be noticed that the \merunetdem{} model has the best PSNR, MSE and SAM, but the \meran{} still has a slightly better SSIM.
One explanation could be that in the \meran{} model, all convolutions are performed in the original resolution, and no downsampling is performed, which might preserve the structural similarity, hence a better SSIM.
The lowest metrics are obtained with the \merunetwosar{} model, which does not use DEM and SAR inputs, showing the benefits of these modalities for the reconstruction.

\begin{table}[H]
\begin{center}
\begin{tabular}{|c|c|c|c|c|}
\hline
Model         & MSE    &  SSIM  & PSNR   & SAM    \\ \hline
\merunetwosar & 324508 & 0.8388 & 24.888 & 0.1595 \\ \hline
\meran        & 277971 & \textbf{0.8656}  & 25.560 & 0.1425 \\ \hline
\merunet      & 261223 & 0.8568 & 25.830 & 0.1448 \\ \hline
\merunetdem   & \textbf{234410} & 0.8645 & \textbf{26.300} & \textbf{0.1401} \\ \hline
\end{tabular}
\caption{Comparison of SSOP networks performed on the \dmer{} dataset}
\label{tab_results_ssop}
\end{center}
\end{table}

\begin{figure}[ht]
\begin{center}
\includegraphics[width=1.0\columnwidth]{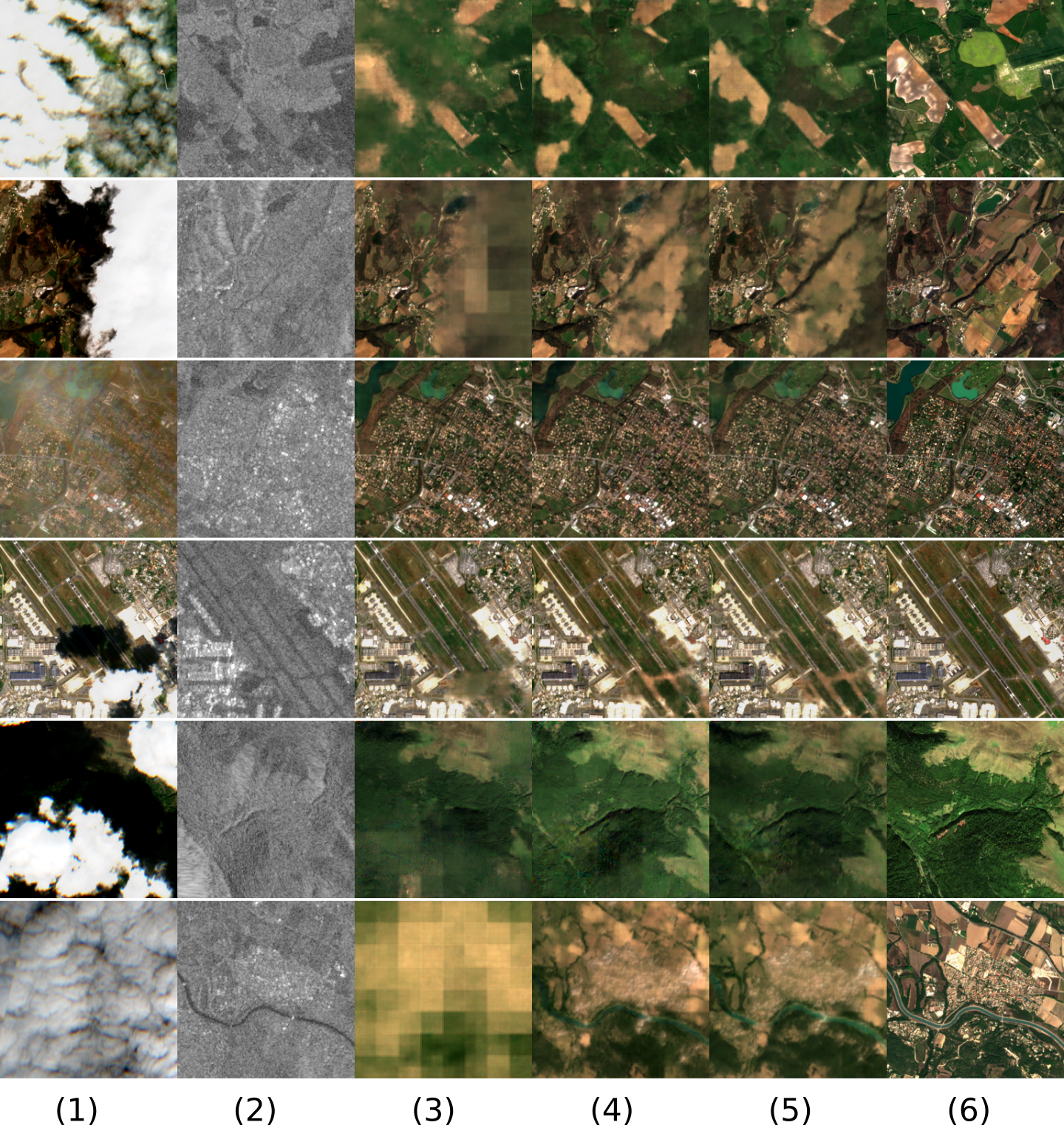}
\caption{From left to right: input cloudy optical image $S2_t$ (1), input SAR image $S1_t$ (2), output $\widehat{S2_t}$ from \merunetwosar{} (3), \merunet{} (4), and \merunetdem{} (5), (6) reference image}
\label{fig_ssop_image_1}
\end{center}
\end{figure}

\begin{figure}[ht]
\begin{center}
\includegraphics[width=0.4\columnwidth]{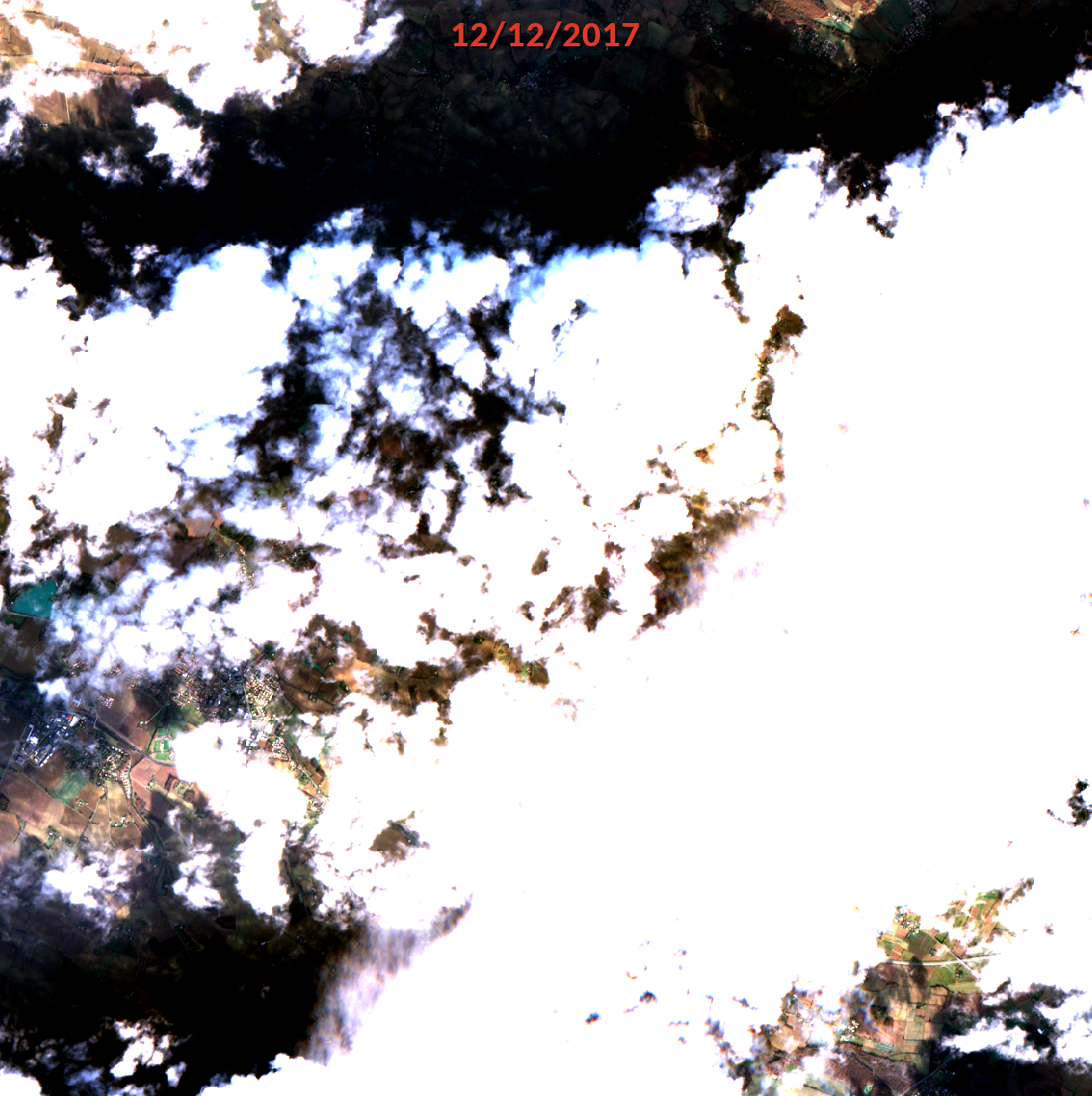}
\includegraphics[width=0.4\columnwidth]{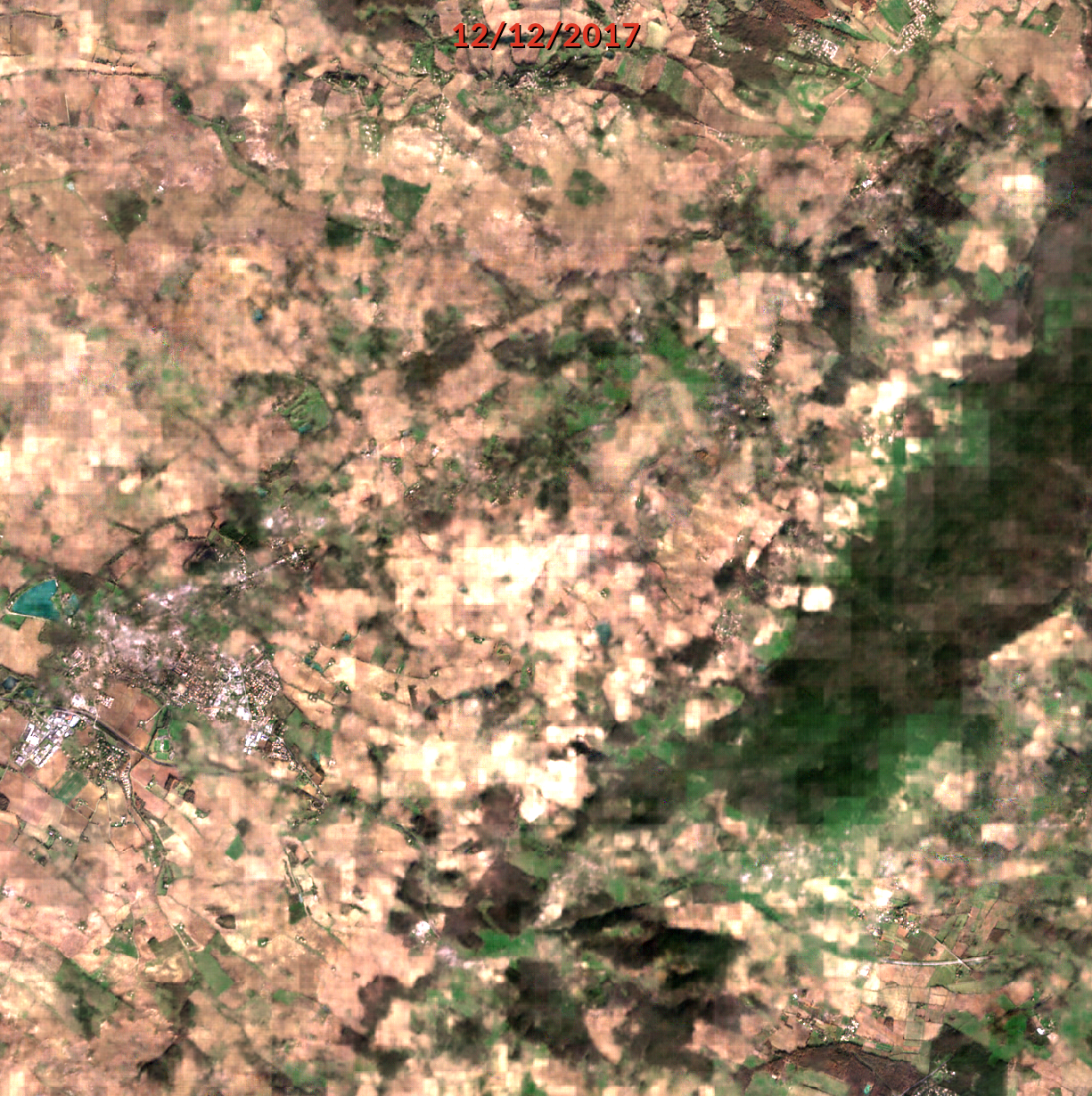}
\includegraphics[width=0.4\columnwidth]{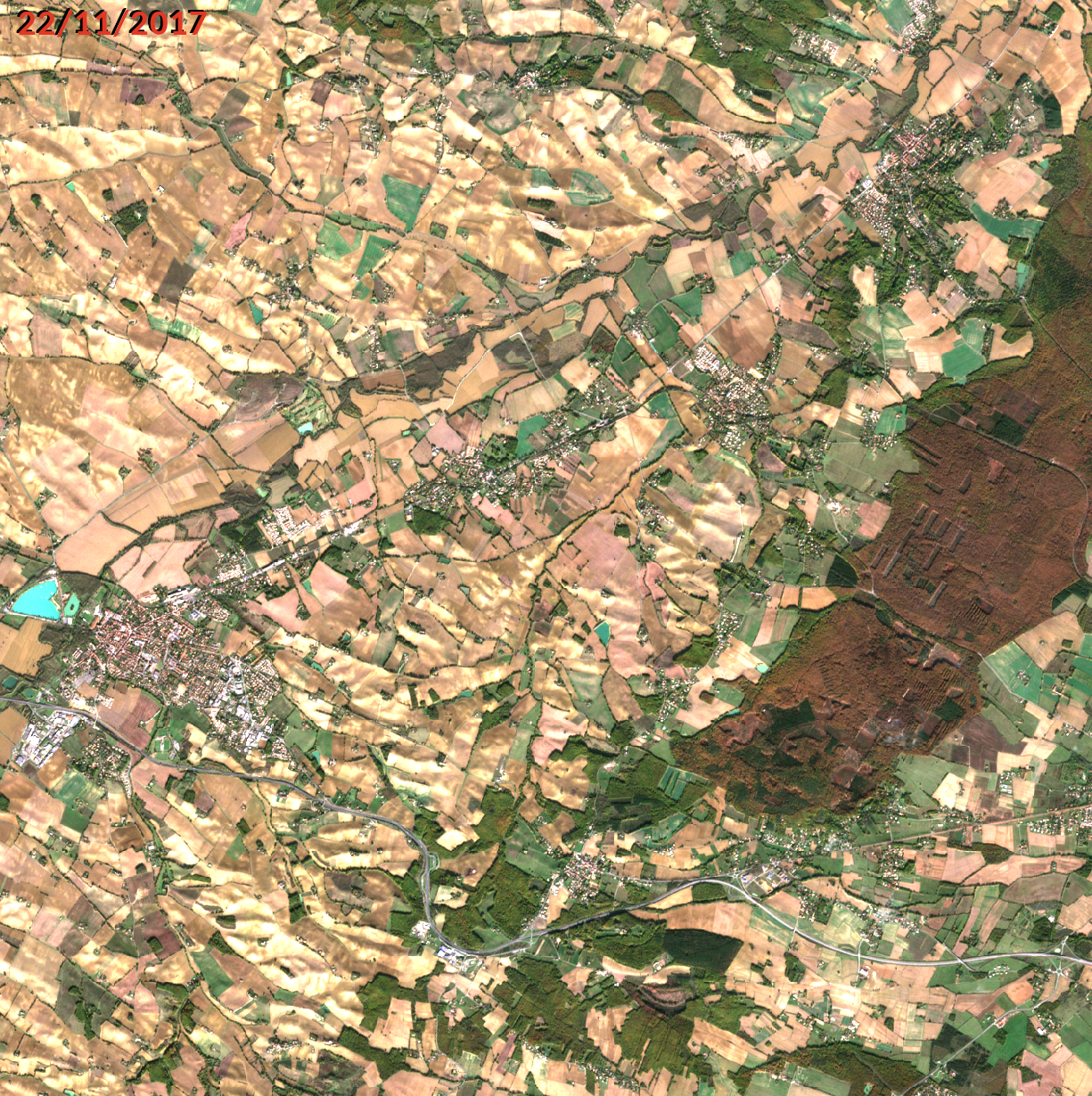}
\caption{Limits of the SSOP networks with thick clouds in optical images. Top left: input cloudy optical image $S2_t$, Top right: SSOP network output $\widehat{S2_t}$. Bottom: the reference image $S2_{t'}$.}
\label{fig_ssop_image_2}
\end{center}
\end{figure}

Figure \ref{fig_ssop_image_1} shows images from the test dataset, processed with the different U-Net based SSOP networks. 
We can visually appreciate the contributions of the input SAR and DEM.
It can be noticed the limits of the method with thick clouds in the optical image, especially for the \merunetwosar{} network that only use the input optical image.
Figure \ref{fig_ssop_image_2} show the limits of the \merunetdem{} network with very thick atmospheric perturbation.

\subsection{Comparison of SSOP and MSOP networks}
\label{sss_res_ssop_msop}

\begin{figure}[ht]
\begin{center}
\includegraphics[width=1.0\columnwidth]{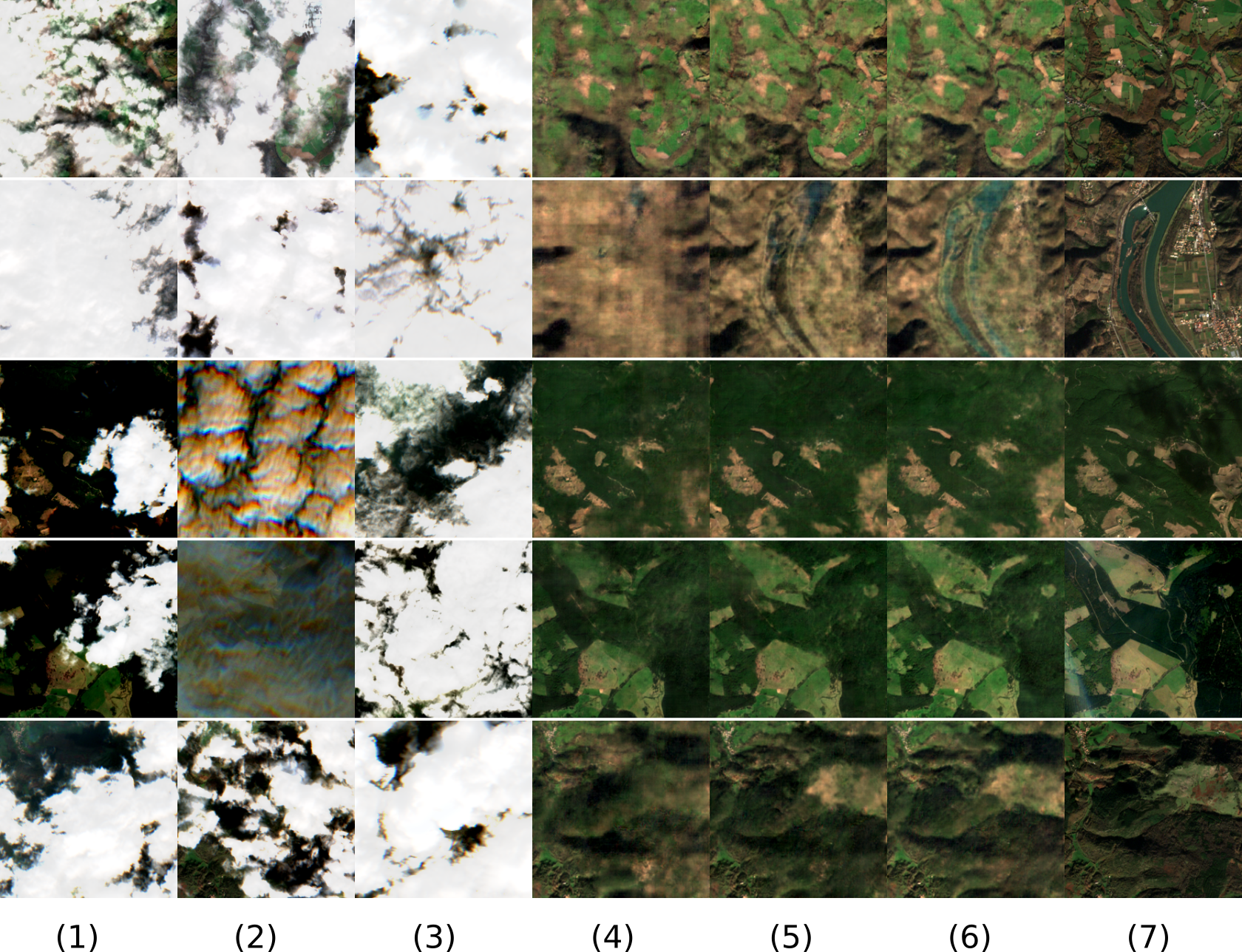}
\caption{From left to right: input images $S2_{t-1}$ (1), $S2_t$ (2), $S2_{t+1}$ (3), output reconstructed optical images $\widehat{S2_t}$ from \scawosar{} (4), \sca{} (5), \scadem{} (6) and the reference image $S2_{t'}$ (7).}
\label{fig_msop_image_1}
\end{center}
\end{figure}

The comparison of approaches that input one or more cloudy images to reconstruct the optical image at $t$, is carried out.
We compare networks that consume different kind of inputs, i.e. one single (S1, S2) pair for SSOP networks versus three pairs of images for MSOP networks.
We recall that, unlike the MSOP networks, $t-1$ and $t+1$ images are not used by the SSOP networks.
We compare the networks on the test dataset detailed in section \ref{sss_al_os21}, since its acquisitions layout fulfills both MSOP and SSOP models validity domains, in particular the maximum SAR-optical temporal gap at $t$.
Evaluation metrics are reported in table \ref{tab_results_ssop_msop}.
While it can be observed the same outcome in the comparison between SSOP networks, these quality metrics differ a bit from the ones presented in table\ref{tab_results_ssop}, since the evaluated samples are just a subset of this last dataset.
The qualitative inspection of the reconstructed images shows that the \scadem{} network produces images better reconstructed than the \sca{} and \scawosar{} networks, especially under thick cloudy areas, highlighting the importance of the SAR and DEM modalities (figure \ref{fig_msop_image_1}).

\begin{table}[H]
\begin{center}
\begin{tabular}{|c|c|c|c|c|}
\hline
Model         & MSE    &  SSIM  & PSNR   & SAM    \\ \hline
\merunetwosar & 324099 & 0.8388 & 24.893 & 0.1595 \\ \hline
\merunet      & 260827 & 0.8567 & 25.836 & 0.1448 \\ \hline
\merunetdem   & 221909 & 0.8583 & 26.538 & 0.1390 \\ \hline
\scawosar     & 141283 & 0.9249 & 28.499 & 0.1128 \\ \hline
\sca          & 138212 & 0.9267 & 28.594 & 0.1111 \\ \hline
\scadem       & \textbf{133061} & \textbf{0.9277} & \textbf{28.759} & \textbf{0.1095} \\ \hline
\end{tabular}
\caption{Comparison between SSOP and MSOP networks, over the \dcrga{} test dataset detailed in section \ref{sss_al_os21}.}
\label{tab_results_ssop_msop}
\end{center}
\end{table}

\subsubsection{Comparison of deep-learning based approaches and gap-filling}
\label{sss_res_msop_gapfil}

\begin{figure}[ht]
\begin{center}
\includegraphics[width=1.0\columnwidth]{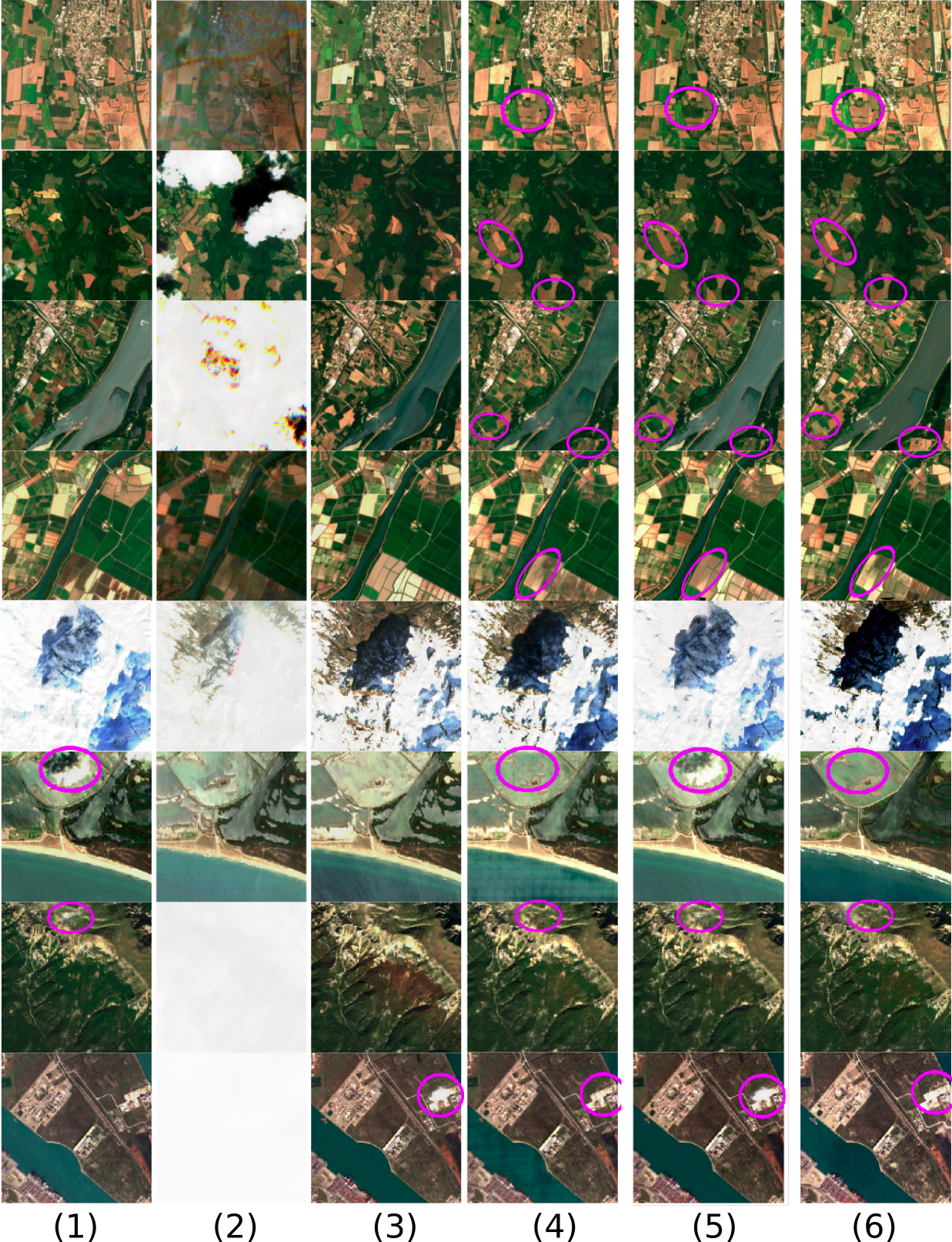}
\caption{From left to right: input images $S2_{t-1}$ (1), $S2_t$ (2), $S2_{t+1}$ (3), output reconstructed optical images $\widehat{S2_t}$ from \scadem{} (4), the Gap-filling (5), and the reference image $S2_{t'}$ (6). In violet are circled details in the reconstructed images that the gap-filling fails to retrieve, or artifacts caused by wrong could masks in input images.}
\label{fig_comp_all}
\end{center}
\end{figure}

In this setup, optical images acquired at $t-1$ and $t+1$ are completely cloud-free, which enables the use of the gap-filling.
Also, the optical image acquired at $t$ is completely covered by clouds or clouds shadows, according to the cloud masks, helping toward a fair comparison between the approach that consume the optical image at $t$ (\sca{}) and the gap-filling.
We perform the comparison of MSOP, SSOP models and the gap-filling using the \dgfa{} test dataset detailed in table \ref{tab_crga_gfa}, which matches the validity range of all approaches. 
Table \ref{tab_results_allmsop} reports the metrics obtained.
We can observe that all the metrics are in favor of the MSOP models.
Also, in this particular use-case, the gap-filling leads to superior results to the SSOP models.
We can notice that all metrics are largely in favor of the \scadem{} model, showing the benefit of the multitemporality, the SAR and the DEM modalities.
Figure \ref{fig_comp_all} shows reconstructed images from the test dataset.
We can notice that the gap-filling fails to retrieve various details in the reconstructed images, like sudden crops changes.
Also, the input images cloud masks are not always exact, and the gap-filling might interpolates polluted images, unlike the MSOP models which are capable of removing those clouds.

\begin{table}[H]
\begin{center}
\begin{tabular}{|c|c|c|c|c|}
\hline
Model         & MSE   &  SSIM  & PSNR & SAM   \\ \hline
\merunetwosar & 239238 & 0.7911 & 26.212 & 0.1847 \\ \hline
\merunet      & 178284 & 0.8187 & 27.489 & 0.1472 \\ \hline
\merunetdem   & 157663 & 0.8264 & 28.023 & 0.1422 \\ \hline
gapfilling    & 79904  & 0.9249 & 30.974 & 0.1021 \\ \hline
\scawosar     & 63097 & 0.9338 & 32.000 & 0.0952 \\ \hline
\sca          & 61016 & 0.9345 & 32.146 & 0.0940 \\ \hline
\scadem       & \textbf{52814} & \textbf{0.9421} & \textbf{32.772} & \textbf{0.0901} \\ \hline
\end{tabular}

\caption{Comparison of the gap-filling and the SSOP and MSOP networks. All approaches are compared on the \dgfa{} test dataset detailed in table \ref{tab_crga_gfa}.}
\label{tab_results_allmsop}
\end{center}
\end{table}

\section{Discussion}
\label{s_discussion}

We have compared various single date SAR/optical networks, with an ablation study to analyze the contribution of the SAR, optical, and DEM inputs.
We have modified the original network from \cite{meraner2020cloud}, which is considerably greedy in term of computational resources, replacing the ResNet backbone with a U-Net backbone.
This has two advantages: first, it is less computationally extensive since convolutions are performed on downsampled features maps.
The processing time is diminished with a factor greater than 30, and leads to similar image quality metrics, with a slightly higher PSNR, but slightly lower SAM and SSIM.
Secondly, using a U-Net backbone instead of a ResNet backbone enables input images at lower resolution that the 10m bands of Sentinel images: we have shown that a 20m spacing DEM can be injected after the first downsampling of the network without prior spatial re-sampling, improving the reconstruction of optical images.
However, we only have trained all single date based networks using the only the \loss{} loss, and future works could investigate other objective formulations.
We have carried out the comparison of single date networks and the multitemporal networks over the \dcrga{} test dataset, which represents the nominal operational context of both networks, e.g. using every available input images, cloudy or not.
Our results shown that the multitemporal networks lead to superior image reconstruction.
We believe that more available input images improves the retrieval of the missing contents of the cloudy optical image at $t$.
The comparison between the deep learning based networks and the gap-filling is performed over the \dgfa{} test dataset, which contains samples where the $t$ optical image patches are covered by clouds at $100\%$.
The gap-filling performs better than the single date network with a significant margin in this particular setup.
However, even though the multitemporal network is not primarily designed for this task, it has outperformed the gap-filling.
Finally our ablation study shows that the SAR and the SAR+DEM contribute in the optical image reconstruction process in both single date based networks, and multitemporal based networks.
For future works, we believe that a further investigation of the SAR signal contribution should be carried out. 
For instance, it could be interesting to study if feeding geometrical information (e.g. local SAR incidence angle) in networks would help, and if physical based SAR pre-processing (i.e despeckeling or target decomposition) benefit the optical image reconstruction task.

\section{Summary and conclusion}

In this paper, we sought to provide a comparison of single date based and multitemporal convolutional networks with the traditional deterministic temporal interpolation between two images.
We have introduced a framework to generate various datasets to train and evaluate various methods for cloudy optical image reconstruction.
Our simple yet convenient method relies on space partitioning data structures indexing the crucial parameters of the remote sensing acquisitions, i.e. how SAR and optical remote sensing images must  be acquired in the datasets, in term of cloud coverage, SAR/optical maximum gap, number and type of acquisition, and relative acquisition time.
We have built several datasets to train single date based networks and multitemporal networks, and to evaluate the different selected approaches, representing various operational contexts.
The studied single date based network take their roots in an existing architecture that uses a ResNet backbone, and we have shown how it could be improved using a U-Net backbone, increasing its training and inference speed and enabling to input additional image of different scale.
We have built a multitemporal network that generalize the single date image reconstruction from three input pairs of images, and which uses the same backbone shared across the inputs.
Our model inputs three cloudy optical and SAR images pairs acquired at dates $t-1$, $t$ and $t+1$, and a DEM.
We have lead the comparison between the single date networks, the multitemporal networks, and the gapfilling in various contexts, showing that the gapfilling performs better than the single date based networks in the context of Sentinel-2 time series.
We have analyzed the contribution of the different kind of inputs, namely optical, SAR and DEM with an ablation study, showing how the reconstructed image benefits from these modalities.
Also, we have shown that, even if the primary design of the multitemporal convolutional network is not focused on image interpolation in temporal domain, it leads to similar even better results than the gap-filling.
However, we should interpret our results carefully regarding the ancillary data available for cloud coverage characterization, since our cloud coverage information per patch depends from it, and the bias it can introduce.
Finally, we lead our study over a small area that do not represents the various atmospheric conditions all over the earth.
With the continuous stream of synchronized SAR and optical acquisitions thanks to the Sentinel constellation, it is expected that future data driven multitemporal models will help to deliver more exploitable data.
Our dataset generation framework and our models are available as open-source software \footnote{http://github.com/cnes/decloud}

\section*{ACKNOWLEDGEMENTS}
\label{ACKNOWLEDGEMENTS}
The authors would like to thank the reviewers for their valuable suggestions.
This work was granted access to the HPC resources of IDRIS under the allocation AD011011608 made by GENCI.

\bibliographystyle{unsrtnat}
\bibliography{paper_arxiv}  

\end{document}